\documentstyle[12pt,psfig,aps]{revtex}

\begin{document}
\pagenumbering{arabic}
\title{\bf Kaon effective mass and energy from  a novel 
chiral SU(3)-symmetric Lagrangian}
\author{Guangjun Mao$^{1}$, P. Papazoglou$^{1}$, S. Hofmann$^{1}$, 
 S. Schramm$^{2}$,
H.~St\"{o}cker$^{1}$, and W.~Greiner$^{1}$}  
\address{$^{1}$Institut f\"{u}r Theoretische Physik der
J. W. Goethe-Universit\"{a}t \\
Postfach 11 19 32,  D-60054 Frankfurt am Main, Germany \\
$^{2}$GSI Darmstadt, Postfach 11 05 52, D-64220 Darmstadt, Germany}
\date{\today}
\maketitle
\begin{abstract}
\begin{sloppypar}
A new chiral SU(3) Lagrangian is proposed 
 to describe the properties of
kaons and antikaons in the nuclear medium, 
the ground state of  dense matter and the
kaon-nuclear interactions consistently. 
 The saturation properties of nuclear matter are reproduced
as well as the results
of the Dirac-Br\"{u}ckner theory. After taking into account the coupling 
between the omega meson and the kaon, we obtain similar results for the 
effective kaon and antikaon energies as calculated in the one-boson-exchange 
model while in our model the parameters of the kaon-nuclear interactions
are constrained by the SU(3) chiral symmetry. 
\vspace{-0.3cm}

\end{sloppypar}
\bigskip
\noindent {\bf PACS} number(s): 14.40.Aq, 12.39.Fe, 21.30.Fe
\end{abstract}
\newcounter{cms}
\setlength{\unitlength}{1mm}
\newpage
\begin{sloppypar}
The properties of kaons/antikaons in nuclear and neutron matter have
attracted considerable interest since the pioneering work of Kaplan and
Nelson \cite{Kap86},  who first proposed
the occurrence of kaon condensation at several
times normal nuclear density. 
Considerable theoretical effort 
has  been
devoted to investigate such medium effects on  kaons 
and antikaons in dense 
matter \cite{Mul90,Bro92,Lut94,Sch94,Kno95,Koc94,Waa96,Ose97}. Recent data by 
the Kaos collaboration \cite{Bar97} on $K^{+}$ and $K^{-}$ production in 
relativistic heavy-ion collisions,
which seems to exhibit
 a substantial enhancement of the $K^{-}$ yield, stimulated 
further activity \cite{Li97}. Among the various  models proposed, the chiral   
SU(3) Lagrangian seems to be particularly useful, since the 
 kaon is  essentially a pseudo-Goldstone
boson. 
The effective Lagrangian of chiral perturbation theory
 used in Refs. \cite{Kap86,Bro92}
reads 
 \begin{equation}
{\cal L}_{KN}= - \frac{3i}{8f_{K}^{2}}\bar{\psi}\gamma_{\mu}\psi \bar{K}
 \stackrel{\leftrightarrow}{\partial^{\mu}}K + \frac{\Sigma_{KN}}{f_{K}^{2}}
 \bar{\psi}\psi\bar{K}K.
 \end{equation}
Here the iso-spin dependent terms have been dropped since the 
problem will be discussed in symmetric  nuclear matter. 
The second term of the above equation stems from the
explicit breaking of chiral symmetry. 
Usually, only the linear order of current-quark mass is taken into account,
which leads to $f_{K}=f_{\pi}=93$ MeV \cite{Ber95}; therefore, 
in the following  
 $f_{K}$ in Eq. (1) is replaced by $f_{\pi}$. 
The amplitude of the $KN$ sigma term has not yet been determined unambiguously.
Two different
choices, $\Sigma_{KN} \approx 2m_{\pi}$ and $\Sigma_{KN}=450\pm30$ MeV,
have been proposed,
in accordance with the Bonn model \cite{Mul90} and with 
lattice gauge calculations
\cite{BroRho}, respectively.  
The kaon and antikaon effective-masses and -energies in static nuclear matter
can be easily derived from Eq. (1)
  \begin{eqnarray}
&& m^{*2}_{K} = m_{K}^{2} - \frac{\Sigma_{KN}}{f_{\pi}^{2}}\rho_{S}, \\        
&& \omega_{K,\bar{K}}=  \sqrt{ m^{*2}_{K}
 + \left( \frac{3}{8f_{\pi}^{2}}\rho_{B} \right) ^{2} }
 \pm \frac{3}{8f_{\pi}^{2}}\rho_{B} .
  \end{eqnarray}
The minus sign in Eq. (3) corresponds to the antikaon energy. $\rho_{S}$ and
$\rho_{B}$ are the scalar and the vector (net) baryon density, respectively. 
Since  chiral
perturbation theory has no direct relation to  the ground state properties 
of the dense
matter, one usually uses \cite{Sch94}
the $\rho_{S}$ and $\rho_{B}$ vacuous as calculated by the
relativistic mean-field theory of Walecka model \cite{Ser86}, i.e.,  a
non-chiral Lagrangian, or simply set $\rho_{S}=\rho_{B}$ \cite{Kap86}, 
in order to evaluate Eqs. (2) and (3). Therefore, the self-consistency
of the theory is lost.

A different approach, based on the chiral SU(3) Lagrangian,
 is the coupled channel
model \cite{Waa96,Ose97}. By including the $\Lambda(1405)$
as a $K^{-}p$ quasi-bound state, the model gives a strong, non-linear density
dependence of the $K^{-}$ potential, which changes sign from positive to 
negative values at low density (around $0.1\rho_{0}$, where 
$\rho_{0}$ is the ground-state
nuclear matter density). This trick allows for an      
attractive potential of antikaons without violating the low density theorems.
Once the density exceeds $0.2\rho_{0}$, the repulsive effect of the 
$\Lambda(1405)$ is neglected - it is predicted that the $\Lambda(1405)$ 
melts in the dense medium, in analogy to a Mott phase transition (for an 
alternative analysis see, however, Ref. \cite{Lut98}).
However, this model does not take into account
 the saturation properties of the system.
Both pion, nucleon and hyperon masses are kept constant in these calculations,
at different densities. 
Up to now, a  consistent calculation based on a chiral Lagrangian, which can 
simultaneously describe both 
the kaon-nuclear interactions and the ground state of the dense matter,
has not been performed yet. 
 
 \end{sloppypar}
 \begin{sloppypar}
This is the aim of the present work.
It addresses these problems in a novel chiral SU(3)-symmetric Lagrangian
\cite{Pap98}. 
In addition to the ground-state saturation properties of 
nuclear matter, the whole density dependence of the mean fields as  
predicted by the Dirac-Br\"{u}ckner theory \cite{Bro90} are considered 
as a further 
constraint to the model. This will turn out to be rather
 important for the investigation of the kaon
and antikaon properties at higher densities.
The chiral SU(3) Lagrangian reads
 \begin{equation}
 {\cal L} = {\cal L}_{kin} + \sum_{W=X,V,u,\Gamma,A}{\cal L}_{BW}
 + {\cal L}_{vec} + {\cal L}_{0} + {\cal L}_{SB}.
 \end{equation}
Here ${\cal L}_{kin}$ is the kinetic energy term, ${\cal L}_{BW}$ gives
the various interaction terms of the different baryons with 4 lowest (spin-0 and
spin-1) meson-multiples
 and with the photons. ${\cal L}_{vec}$ generates the 
 vector meson-masses through interactions
 with the spin-0 mesons, and ${\cal L}_{0}$
 gives the meson-meson interaction potentials which includes the spontaneous  
breaking of chiral symmetry and trace anomaly. Finally, ${\cal L}_{SB}$ introduces an explicit
symmetry breaking of U(1)$_{A}$, SU(3)$_{V}$, and the chiral symmetry.
The main feature of the model is that the baryon masses are generated by the
scalar condensates while their splitting is realized through SU(3) symmetry
breaking for these condensates.
The model is described in  \cite{Pap98}.
Considering  SU(3) generators up to quadratic order, the Lagrangian for
nuclear matter reads 
 \begin{eqnarray}
&& {\cal L}_{kin} = i \bar{\psi}\gamma_{\mu}\partial^{\mu}\psi + \frac{1}{2}
 \partial_{\mu}\sigma\partial^{\mu}\sigma + \frac{1}{2}\partial_{\mu}\zeta
 \partial^{\mu}\zeta + \frac{1}{2}\partial_{\mu}\chi\partial^{\mu}\chi
  - \frac{1}{4}\omega_{\mu\nu}\omega^{\mu\nu}, \\
&& {\cal L}_{NX} + {\cal L}_{NV} = - \bar{\psi} \left[ {\rm g}_{N\omega}
 \gamma_{\mu}\omega^{\mu} + m_{N} \right] \psi, \\
&& {\cal L}_{vec} = \frac{1}{2}m_{\omega}^{2} \frac{\chi^{2}}{\chi_{0}^{2}}
 \omega_{\mu}\omega^{\mu} + {\rm g}_{4}^{4}\left( 
 \omega_{\mu}\omega^{\mu} \right)^{2}, \\
&& {\cal L}_{0} = - \frac{1}{2}k_{0}\chi^{2}\left( \sigma^{2} + \zeta^{2}
 \right) + k_{1} \left( \sigma^{2} + \zeta^{2} \right)^{2} 
 + k_{2} \left( \frac{\sigma^{4}}{2} + \zeta^{4} \right) 
 + k_{3} \chi \sigma^{2} \zeta \nonumber \\
 && \;\;\;\;\;\;\;\;\;
 - k_{4}\chi^{4} + \frac{1}{4}\chi^{4} {\rm ln} \frac{\chi^{4}}{\chi_{0}^{4}}
 + \frac{\delta}{3}\chi^{4}{\rm ln} 
 \frac{\sigma^{2}\zeta}{\sigma_{0}^{2}\zeta_{0}},\\
&& {\cal L}_{SB} = - \left( \frac{\chi}{\chi_{0}}\right) ^{2}
 \left( f_{1}\sigma + f_{2} \zeta\right), 
 \end{eqnarray}
and the kaon interaction is described by
\begin{equation}
{\cal L}_{KN} = - \frac{3i}{8f_{K}^{2}}\bar{\psi}\gamma_{\mu}\psi
\bar{K}\stackrel{\leftrightarrow}{\partial^{\mu}}K + \frac{m_{K}^{2}}
{2f_{K}} \left( \sigma + \sqrt{2}\zeta \right) \bar{K}K
-i{\rm g}_{\omega K}\bar{K}\stackrel{\leftrightarrow}{\partial^{\mu}}K
\omega_{\mu}.
\end{equation}
Here $\sigma$, $\omega$ are the scalar and vector field and 
$\zeta$, $\chi$ are the strange scalar field and the gluon field, 
respectively; $f_{1}=m_{\pi}^{2}f_{\pi}$, $f_{2}=\sqrt{2}m_{K}^{2}f_{K}
 - \frac{1}{\sqrt{2}}m_{\pi}^{2}f_{\pi}$, $m_{N}=m_{0} - \frac{1}{3}{\rm g}_{8}
 ^{S}(4\alpha -1)(\sqrt{2}\zeta - \sigma)$, $m_{0}={\rm g}_{1}^{S}(\sqrt{\frac
 {2}{3}}\sigma + \sqrt{\frac{1}{3}}\zeta)$. 
The omega-kaon coupling is introduced through considering the vector field
as a gauge field.
The vacuum condensates of the scalar
fields $\sigma_{0}$ and $\zeta_{0}$ generate the masses of the various hadrons.
They are constrained by the pion and kaon decay constants: 
$f_{\pi}=-\sigma_{0}$, $f_{K}=-(\sigma_{0}+ \sqrt{2}\zeta_{0})/2$. If the 
equality $\sigma_{0}=\sqrt{2}\zeta_{0}$ is satisfied, the model regains the
SU(3) symmetry. 
The parameters of the model are $\sigma_{0}$, $\zeta_{0}$, $\chi_{0}$,
${\rm g}_{N\omega}$, 
$\sum_{i=0}^{4}k_{i}$, $\delta$, ${\rm g}_{4}$ and ${\rm g}_{1}^{S}$,
${\rm g}_{8}^{S}$, $\alpha$.
Ten of these are determined by the SU(3) vacuum
and the 18+8 hadron masses,
 three parameters, i.e., the vector coupling constant ${\rm g}_{N\omega}$ 
 and ${\rm g}_{4}$ plus the "gluon condensate" $\chi_{0}$
 are used to fit the saturation properties of nuclear 
matter. This yields two sets of parameters denoted as C1 and C2.
These two parameters sets differ in the strange condensate $\zeta$. C1 allows
for an explicit $\zeta$-dependence of $m_{N}$, while C2 excludes such a
dependence.
 The corresponding saturation properties are
(1) C1: $m^{*}/m_{N}=0.612$, $E/A(\rho_{0})=-15.99$ MeV, $K=276.3$ MeV;
(2) C2: $m^{*}/m_{N}=0.641$, $E/A(\rho_{0})=-15.93$ MeV, $K=266.2$ MeV.
Both C1 and C2 have a saturation density $\rho_{0}=0.15$ $fm^{-3}$ and
$f_{\pi}=93.3$ MeV, $f_{K}=122$ MeV, $m_{\pi}=139$ MeV, $m_{K}=498$ MeV.
The parameters of the kaon-nuclear interactions,  
 Eq. (10),
are constrained by the chiral Lagrangian itself. 
${\rm g}_{\omega K}={\rm g}_{\rho\pi\pi}f_{\pi}^{2}/2f_{K}^{2}$, and 
${\rm g}_{\rho\pi\pi}=6.05$ from the $\rho^{0} \rightarrow \pi^{+}\pi^{-}$
decay \cite{PDG}.
The results do not depend 
on the $KN$ sigma term.  
 $\Sigma_{KN}$ is computed in the present model \cite{prep}.

A field shift to new variables, $\phi$ and $\xi$, is performed 
($\sigma=\sigma_{0} - \phi$, $\zeta = 
\zeta_{0} - \xi$) and $\chi=\chi_{0}$ is set (the variation of the gluon
condensate in the nuclear medium is negligible \cite{Pap98}). 
Then the following field equations of the 
scalar and vector mesons in static nuclear matter are obtained 
after some straightforward algebra
  \begin{eqnarray}
&& \left( k_{0}\chi_{0}^{2} -12 k_{1}\sigma_{0}^{2} - 4k_{1}\zeta_{0}^{2}
 -6k_{2}\sigma_{0}^{2}-2k_{3}\chi_{0}\zeta_{0} \right) \phi 
 + \left( 12k_{1}\sigma_{0} +6k_{2}\sigma_{0} \right) \phi^{2} \nonumber \\
&& -\left( 4k_{1} +2k_{2}\right) \phi^{3} + \frac{2\delta}{3}\chi_{0}^{4}
 \frac{1}{\sigma_{0} -\phi} -\frac{2\delta}{3\sigma_{0}}\chi_{0}^{4}
 -4k_{1}\phi\xi^{2} +\left(8k_{1}\zeta_{0}+2k_{3}\chi_{0} \right)\phi\xi
 \nonumber \\
&& +4k_{1}\sigma_{0}\xi^{2} -\left(8k_{1}\sigma_{0}\zeta_{0} +2k_{3}\chi_{0}
 \sigma_{0} \right) \xi =-{\rm g}_{N\sigma}\rho_{S} , \\
&& \left( k_{0}\chi_{0}^{2} -12k_{1}\zeta_{0}^{2} -4k_{1}\sigma_{0}^{2}
-12k_{2}\zeta_{0}^{2} \right) \xi + 12\left( k_{1} +k_{2} \right) \zeta_{0}
 \xi^{2} -4\left( k_{1} + k_{2} \right) \xi^{3} \nonumber \\
&& + \frac{\delta}{3}\chi_{0}^{4}
 \frac{1}{\zeta_{0} - \xi} -\frac{\delta}{3\zeta_{0}}\chi_{0}^{4} 
-4k_{1}\phi^{2}\xi +\left( 4k_{1}\zeta_{0} +k_{3}\chi_{0} \right)\phi^{2}
+8k_{1}\sigma_{0}\phi\xi \nonumber \\ 
&& -\left( 8k_{1}\sigma_{0}\zeta_{0} +2k_{3}\chi_{0}
 \sigma_{0} \right) \phi = -{\rm g}_{N\zeta}\rho_{S} , \\
&& m_{\omega}^{2}\omega + 4{\rm g}_{4}^{4}\omega^{3} ={\rm g}_{N\omega}
 \rho_{B}.
  \end{eqnarray}
The effective-mass and -energy of the kaon $K$ 
and the antikaon $\bar{K}$ are given by
\begin{eqnarray}
&& m^{*2}_{K} = m_{K}^{2} + \frac{m_{K}^{2}}{2f_{K}}\phi
+\frac{m_{K}^{2}}{\sqrt{2}f_{K}}\xi , \\
&& \omega_{K,\bar{K}}=  \sqrt{ m^{*2}_{K}
+ \left( \frac{3}{8f_{K}^{2}}\rho_{B} + {\rm g}_{\omega K}\omega_{0}\right)
^{2} }
\pm \left( \frac{3}{8f_{K}^{2}}\rho_{B} + {\rm g}_{\omega K}\omega_{0}
\right).
\end{eqnarray}
Here the coupling strengths ${\rm g}_{N\sigma}$ and ${\rm g}_{N\zeta}$
are given by   
 \begin{eqnarray}
&& {\rm g}_{N\sigma}=-\left[ \sqrt{\frac{2}{3}}{\rm g}_{1}^{S}
 +\frac{1}{3}{\rm g}_{8}^{S} \left( 4\alpha -1\right) \right], \\
&& {\rm g}_{N\zeta}=-\left[ \sqrt{\frac{1}{3}}{\rm g}_{1}^{S}
 -\frac{\sqrt{2}}{3}{\rm g}_{8}^{S} \left( 4\alpha -1\right) \right]. 
 \end{eqnarray}
In principle, one can implement additional terms on the Lagrangian  
with more than one time derivate acting on the kaon field (i.e., the
so-called off-shell terms), for example,
$ \sim Tr (u_{\mu}u^{\mu}\bar{B}B)$. 
These terms are  not included in the
present work. 

 \end{sloppypar}
 \begin{sloppypar}
Fig.~1 displays the binding energy of the system (a) and the scalar and vector
potentials of the nucleons (b) as a function of 
the Fermi momentum. The results of the
chiral effective Lagrangian with the C1 and C2 parameters as well as the 
relativistic mean field theory with the linear \cite{Ser86} and non-linear
(TM1)  \cite{Sug94} parameterization are 
compared with the prediction of the Dirac-Br\"{u}ckner G-matrix 
theory \cite{Bro90}.
It can be seen that the results of
the linear Walecka model deviate from the Dirac-Br\"{u}ckner theory evidently.  
However, C1, C2 and TM1 can  reproduce the results of 
the relativistic
G-matrix theory nearly perfectly up to the normal density. This explains why
the description of finite nuclei is rather convincing in these effective models 
\cite{Sug94,Pap98}. At higher densities, the results of C1 and C2 are  
 closer to the prediction of the Dirac-Br\"{u}ckner theory
than TM1. Both C1 and C2 follow the results of the G-matrix calculations closely
 up to four times normal density although the parameters have not been fitted
to the results of Ref. \cite{Bro90}. Therefore, the novel chiral  SU(3)
 Lagrangian \cite{Pap98} yields reasonable results for dense matter
 $\rho \leq 4\rho_{0}$.
These results  provide a sound self-consistent basis for the
investigation of kaon and antikaon properties in the nuclear medium. 

Fig.~2(a) depicts the kaon and antikaon effective-masses  as calculated
from this SU(3) chiral model with the parameter set C2 (the results of parameter
set C1 are nearly indistinguishable from that of C2). 
The results of the Kaplan-Nelson model are
also plotted for two different $KN$ sigma terms,  $\Sigma_{KN}=2m_{\pi}$ and 
$\Sigma_{KN}=450$ MeV  with the mean fields  provided by Eqs. (11)-(13) 
(parameter set C2). The density dependence of the scalar fields as computed 
with the
parameter set C2 is given in Fig.~2(b).
From now on, let us denote the Lagrangian of Eq. (1)  
model I and that of Eq. (4) model II, respectively. 
For model I, 
the effective mass of the kaon decreases nearly linearly with increasing  
density, with a slope determined by  $\Sigma_{KN}$. 
For model II  $m^{*}_{K}$
first decreases with increasing density, but then it approaches  
saturation, consistent with the finding of the one-boson-exchange (OBE) model
 \cite{Sch94}. One can easily understand the different behavior of two models
from Eqs. (2) and (14): In Eq. (2) the $m^{*}_{K}$ depends linearly  
on $\rho_{S}$ (which is, in turn, approximately equal to $\rho_{B}$).
In Eq. (14),  the $m_{K}^{*}$ is related to the
scalar fields. To obtain the Dirac-Br\"{u}ckner results displayed in  Fig.~1,
 a highly non-linear relation
between the $\phi$, $\xi$ and the $\rho_{S}$ is asked for, 
as can be seen from Eqs. (11) and (12).
Fig.~2(b)  shows  the saturation of the scalar fields 
at high densities.
Consequently, the $m^{*}_{K}$ saturates after $\rho_{B} \approx
2\rho_{0}$,  it never tends to zero.  
 Neither kaon nor antikaon condensation occur
in the chiral SU(3)  model. 
In fact,  antikaon condensation  could occur only if 
the condensates of the scalar fields, $\sigma$ and $\zeta$, would 
vanish in the medium. This would constitute  a
chiral phase transition. Note that  antikaon condensation may occur in model
I depending on the values of $\Sigma_{KN}$  
 as well as on the  mean field \cite{Sch94}
used in the calculations.

Fig.~3 compares the chiral SU(3) model calculations (with the parameter set C2) 
for the kaon and antikaon effective-energies, with (solid lines) and without
(dot-dashed lines) the omega-kaon coupling,  
  with the results of other models 
\cite{Kap86,Sch94}. 
This is done up to $3\rho_{0}$, where all 
these models should be least unreliable (than in the regime of 
extremely high densities). 
In additional, an "empirical" kaon dispersion
relation  \cite{Li97} is also presented - it resulted 
from "fitting" the Kaos data
\cite{Bar97} of $K^{+}$ and $K^{-}$ production in heavy-ion collisions 
by means of a relativistic 
transport model \cite{Ko96} by adjusting the real part of the kaon and antikaon
optical potential, but ignoring the imaginary part (i.e., the 
in-medium scattering cross sections and hyperon resonances).
It can be seen that without the $\omega - K$ coupling our model
predicts a rather weak potential for the antikaon compared to the
predictions of other models. After introducing the
$\omega - K$ coupling, the calculated effective energies for kaon and
antikaon are quite similar as obtained in the one-boson-exchange model
\cite{Sch94}. The effective kaon and antikaon energies are directly related
to their optical potentials. The chiral SU(3) model gives $U_{opt}^{\bar{K}}
= -98.7$ MeV and $U_{opt}^{K} = 21.5$ MeV at normal density.
The predicted $K^{-}$
optical potential is in accordance with the results of model I and OBE
\cite{Sch94} (between $-70$ and $-100$ MeV).
It is close to the empirical value obtained by the standard fit
of $K^{-}$-atomic data, which gives $U_{opt}^{\bar{K}}=-85$ MeV, but much
weaker as compared to the non-linear fit, which gives
$U_{opt}^{\bar{K}}=-200
\pm 20$ MeV \cite{Fri93}. 
At present, there exist no firmly established
empirical value for the $U_{opt}^{K}$. An estimate
based on  impulse approximation \cite{Lut94,Sch94}
gives $U_{opt}^{K} \approx 29$ MeV. Our results are in agreement with it.

In summary, we have employed a recent developed
  chiral SU(3) Lagrangian to investigate the
properties of kaons and antikaons in the nuclear medium. The kaon-nuclear 
interactions and the ground state of the dense matter are described 
consistently for the first time within a chiral approach.
The parameters of the kaon-nuclear interactions are constrained by the SU(3)
chiral symmetry. 
The saturation properties of nuclear matter 
as well as the results of the 
Dirac-Br\"{u}ckner theory are well reproduced.
Due to the highly non-linear kaon interaction with respect to the 
 density, the kaon/antikaon effective mass is changed only 
moderately in the nuclear medium. After introducing the coupling between the
omega meson and the kaon, we obtain similar results for the kaon and
antikaon effective energies as calculated in the one-boson-exchange model,
i.e., the kaon feels a weak repulsive potential while the antikaon suffers
a strong attractive potential.  

 \end{sloppypar}
 \vspace{0.5cm}
 \begin{sloppypar}
 We would like to  thank  D.~Zschiesche and H.~Weber  
 for fruitful discussions. We thank J.~Schaffner for a careful reading of
the preliminary version of the manuscript and pointing out the importance of 
including the omega-kaon coupling.
 G.~Mao is  grateful to the Alexander von
Humboldt-Stiftung for financial support and to the people at the
Institut f\"{u}r
Theoretische Physik der J.~W.~Goethe Universit\"{a}t for their hospitality.
This work was supported by DFG-Graduiertenkolleg Theoretische \& Experimentelle
Schwerionenphysik, GSI, BMBF, DFG and A.v.Humboldt-Stiftung.

 \end{sloppypar}
 
 \newpage
  \vspace{0.5cm}
 \begin{figure}[htbp]
 \hskip  1.0cm \psfig{file=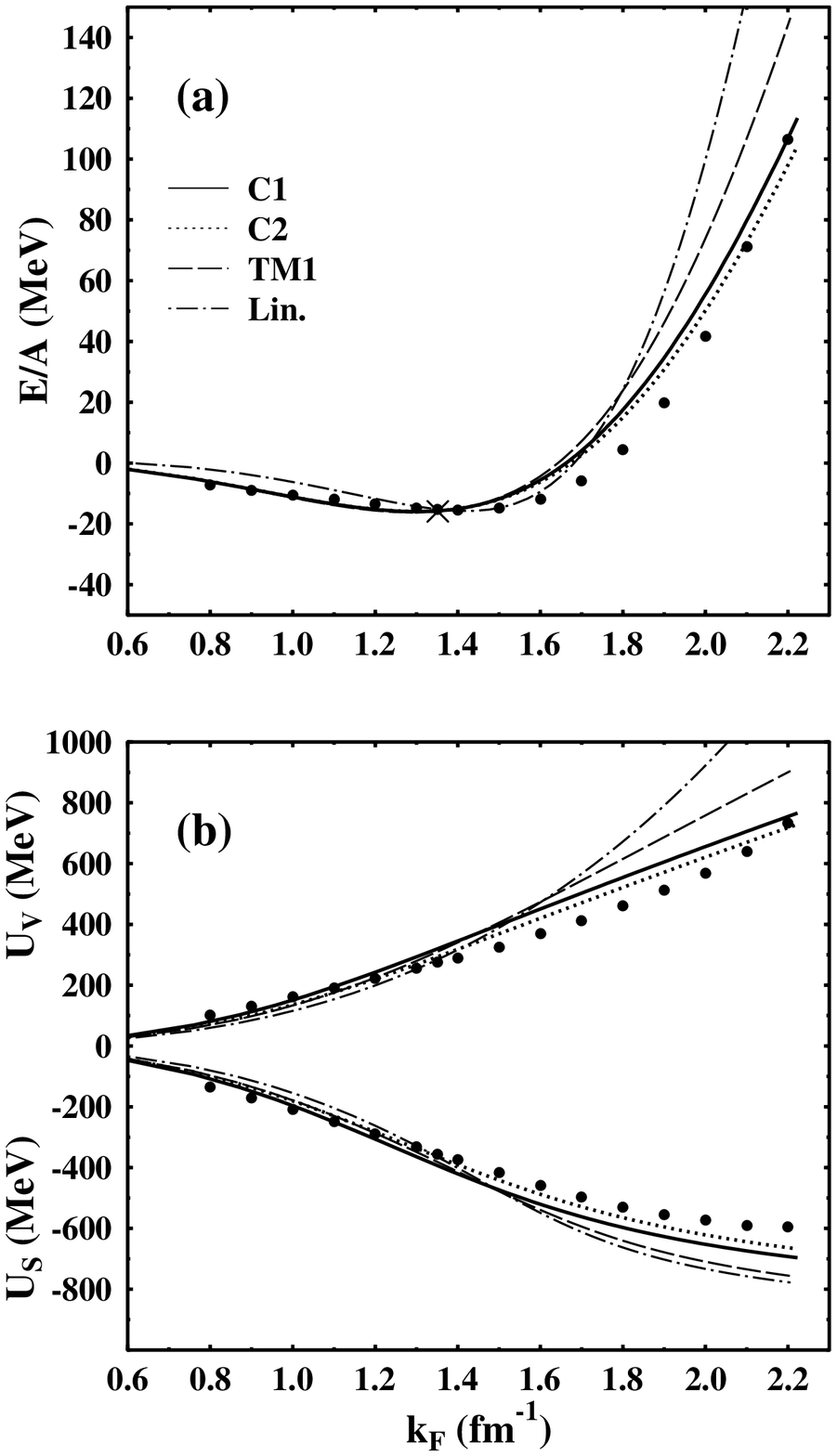,width=11cm,height=17cm,angle=0}
 \vspace{-1cm}
 \caption{(a) The binding energy and (b) the scalar and vector potentials
of the nucleons calculated with the chiral Lagrangian (C1, C2) 
and the linear [14] and non-linear (parameter set TM1 [20])
 relativistic mean field
theory. 
The dots are the results of the Dirac-Br\"{u}ckner theory of 
Ref. [17].  The cross in (a) denotes the empirical value for
nuclear matter saturation ($E/A=-16\pm 1$ MeV, $k_{F}=1.35\pm 0.05$ $fm^{-1}$).
 The scalar and vector potential at saturation point ($k_{F}=1.305$ $fm^{-1}$)
calculated with the chiral Lagrangian are  $-364.3$ MeV, $293.3$ MeV for C1 and
$-337.1$ MeV, $268.4$ MeV for C2, respectively.}
 \vspace{-7cm}
\end{figure}
 \newpage
 \begin{figure}[htbp]
  \vspace{0.0cm}
 \hskip   1.0cm \psfig{file=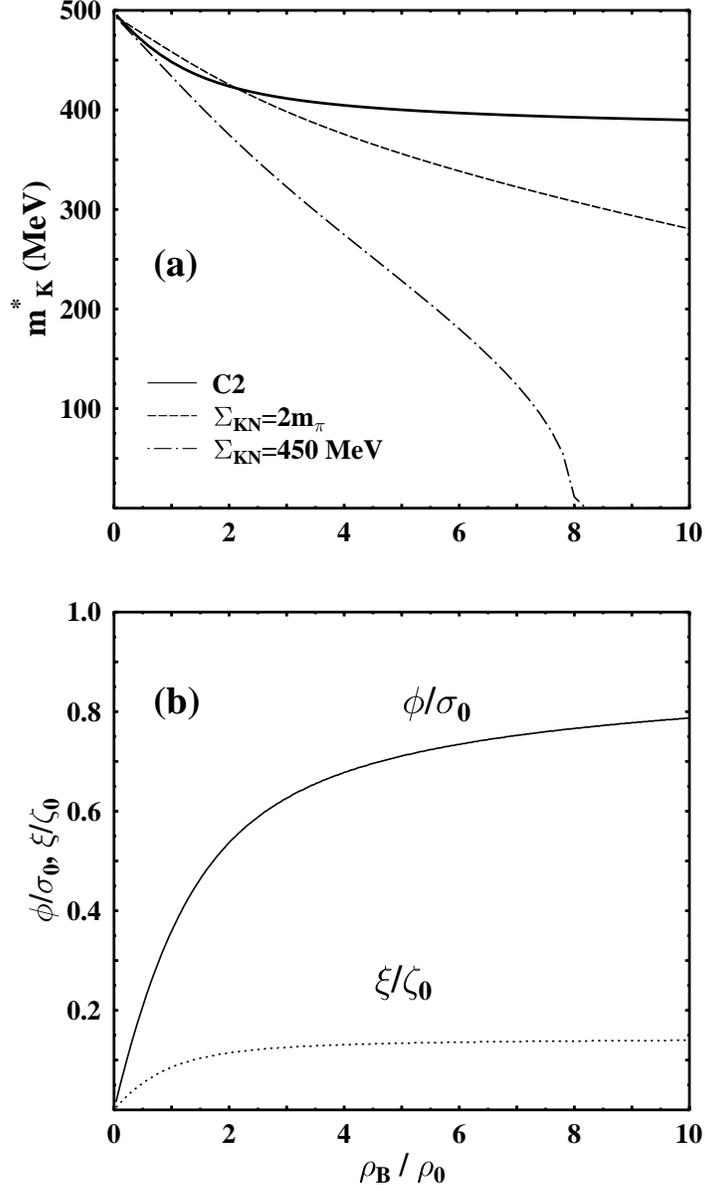,width=11.cm,height=19cm,angle=0}
 \vspace{-1cm}
 \caption{(a) The effective masses of kaons and antikaons  in
 nuclear matter versus the baryon density at T=0. The solid  curve represents
 the results of this work with the parameter set C2. The 
dashed and dot-dashed curves are calculated from Eqs. (2) and (3) 
with the $\rho_{S}$-
and $\rho_{B}$-values as  provided by Eqs. (11) - (13) (parameter set C2).  
(b) The density dependence of the scalar fields as shown for  
the parameter set C2.}
 \vspace{-7cm}
\end{figure}
 \newpage
 \begin{figure}[htbp]
  \vspace{0.0cm}
 \hskip  -1.5cm \psfig{file=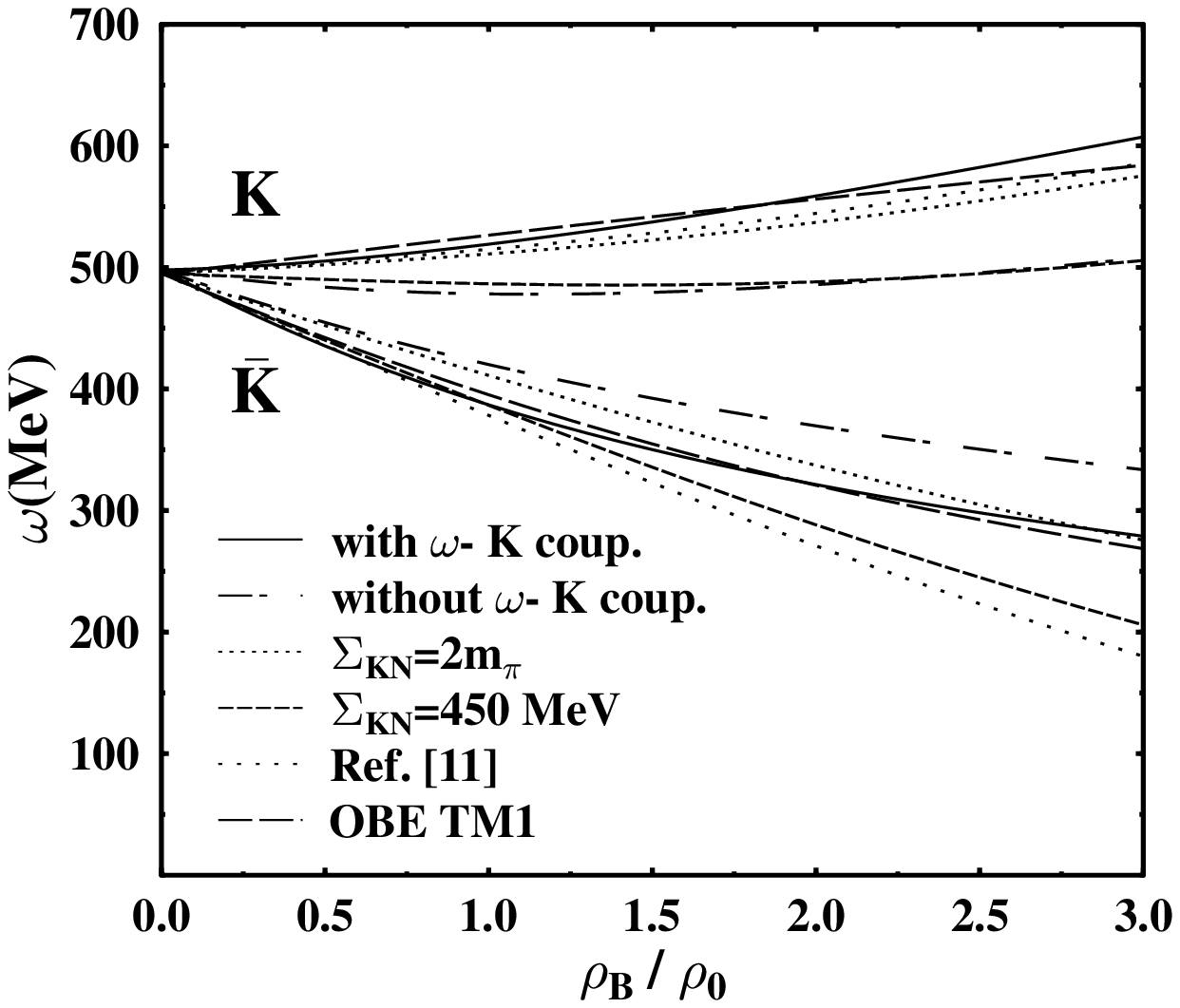,width=14.cm,height=16cm,angle=-90}
 \caption{The energies of kaons and antikaons as a function of the density.
The solid and dot-dashed lines represent the results of this work with and 
without the $\omega - K$ coupling, respectively. The short-dotted and 
short-dashed lines are calculated with the chiral perturbation theory with 
the different $KN$ sigma terms. The long-dotted lines are the results of 
Ref. [11]. The long-dashed lines depict the results of the one-boson-exchange 
model with the TM1 parameter set.}
\end{figure}

\newpage
\begin{thebibliography}{250}
\bibitem{Kap86}
   D.B.~Kaplan and A.E.~Nelson, 
   Phys. Lett. {\bf B175}, 57 (1986);
   A.E.~Nelson and D.B.~Kaplan, 
   Phys. Lett. {\bf B192}, 193 (1987).
\bibitem{Mul90}
   A.~M\"{u}ller-Groeling, K.~Holinde, and J.~Speth,
   Nucl. Phys. {\bf A513}, 557 (1990).
\bibitem{Bro92}
   G.E.~Brown, C.M.~Ko, and K.~Kubodera, 
   Z. Pyhs. {\bf A341}, 301 (1992);
   G.E.~Brown, C.H.~Lee, M.~Rho, and V.~Thorsson,
   Nucl. Phys. {\bf A567}, 937 (1994).
\bibitem{Lut94}
   M.~Lutz, A.~Steiner, and W. ~Weise,
   Nucl. Phys. {\bf A574}, 755 (1994).
\bibitem{Sch94}
   J.~Schaffner, A.~Gal, I.N.~Mishustin, H.~St\"{o}cker, and W. ~Greiner,
   Phys. Lett. {\bf B334}, 268 (1994);
   J.~Schaffner and I.N.~Mishustin,
   Phys. Rev. {\bf C53}, 1416 (1996);
   J.~Schaffner, J.~Bondorf, and I.N.~Mishustin,
   Nucl. Phys. {\bf A625}, 325 (1997).
\bibitem{Kno95}
   R.~Knorren, M.~Prakash, and P.J.~Ellis,
   Phys. Rev. {\bf C52}, 3470 (1995).
\bibitem{Koc94}
   V.~Koch
   Phys. Lett. {\bf B337}, 7 (1994).
\bibitem{Waa96}
   T.~Waas, N.~Kaiser, W.~Weise,
   Phys. Lett. {\bf B365}, 12 (1996);
   Phys. Lett. {\bf B379}, 34 (1996);
   N.~Kaiser, P.B.~Siegel, and W.~Weise,
   Nucl. Phys. {\bf A594}, 325 (1995).
\bibitem{Ose97}
   E.~Oset and A.~Ramos,
   nucl-th/9711022.
\bibitem{Bar97}
   R.~Barth, P.~Senger, W.~Ahner et al., 
   Phys. Rev. Lett. {\bf 78}, 4007 (1997).
\bibitem{Li97}
   G.Q.~Li, C.H.~Lee, and G.E.~Brown,
   Phys. Rev. Lett. {\bf 79}, 5214 (1997).
\bibitem{Ber95}
   V.~Bernard, N.~Kaiser, and Ulf-G.~Meissner,
   Int. J. Mod. Phys. {\bf E4}, 193 (1995).
\bibitem{BroRho}
   G.E.~Brown and M.~Rho,
   Phys. Rep. {\bf 269}, 333 (1996).
\bibitem{Ser86}
   B.~D.~Serot and J.~D.~Walecka,
   Adv. Nucl. Phys. {\bf 16}, 1 (1986).
\bibitem{Lut98}
   M.~Lutz,
   Phys. Lett. {\bf B426}, 12 (1998).
\bibitem{Pap98}
   P.~Papazoglou, S.~Schramm, J.~Schaffner-Bielich, H.~St\"{o}cker, 
   and W.~Greiner,
   Phys. Rev. {\bf C57}, 2576 (1998);
   P.~Papazoglou, D.~Zschiesche, S.~Schramm, J.~Schaffner-Bielich,
   H.~St\"{o}cker, and W.~Greiner,
   nucl-th/9806087, Phys. Rev. C in press.
\bibitem{Bro90}
   R.~Brockmann and R.~Machleidt,
   Phys. Rev. {\bf C42}, 1965 (1990)
\bibitem{PDG}
   Particle Data Group, R.M.~Barnett et al.,
   Phys. Rev. {\bf D54}, 1 (1996).   
\bibitem{prep}
   Guangjun~Mao, P.~Papazoglou, S.~Schramm, H.~St\"{o}cker, and W.~Greiner,
   unpublished.
\bibitem{Sug94}
   Y.~Sugahara, H.~Toki,
   Nucl. Phys. {\bf A579}, 557 (1994).
\bibitem{Fri93}
   E.~Friedman, A.~Gal, and C.J.~Batty,
   Phys. Lett. {\bf B308}, 6 (1993);
   Nucl. Phys. {\bf A579}, 518 (1994);
   see Ref. \cite{newdata}.
\bibitem{Ko96}
   C.M.~Ko and G.Q.~Li,
   J. Phys. {\bf G22}, 1673 (1996). 
\bibitem{newdata}
After finishing this manuscript, we received a new fit to  kaonic
atom data, which uses the linear Walecka model for the interior of nuclei, 
and gives $U_{opt}^{\bar{K}}
=-185\pm 15$ MeV: 
E.~Friedman, A.~Gal, and J.~Mares,  
nucl-th/9804072.

\end{thebibliography}
\end{document}